\documentclass{article}
\usepackage[latin1]{inputenc}
\usepackage{amsmath,amsthm,amsfonts}

\newcommand{\N}{\mathbb{N}}
\newcommand{\Sym}[1]{\mathrm{Sym}\left(#1\right)}

\newcommand{\LMlt}{\mathcal{L}}

\newtheorem{definition}{Definition}
\newtheorem{proposition}{Proposition}

\title{A note on conjugacy search and racks}

\author{Juha Partala\\\\
Department of Electrical and Information Engineering,\\
University of Oulu\\
}

\begin{document}

\maketitle

\begin{abstract}
We show that
for every effective left conjugacy closed left quasigroup, 
there is an induced rack that retains the conjugation structure of the left translations.
This means that cryptographic protocols relying on conjugacy search
can be secure only if conjugacy search of left translations is infeasible
in the induced rack.
We note that, in fact, protocols based on conjugacy search could be simply implemented using a rack.
We give an exposition of the Anshel-Anshel-Goldfeld protocol
in such a case.
\end{abstract}


\section{Introduction}
\label{Introduction}

A cryptographic key exchange protocol allows two or more parties to establish a common key
using an insecure channel. The key can be subsequently used for secure transmission.
Security of a key exchange protocol typically relies on a computationally hard problem.
The conjugacy search problem (CSP) was first
suggested for key exchange in the pioneering work by Anshel et al.~\cite{Anshel_1999,Anshel_2001}
and Ko et al.~\cite{Ko_2000}.
The CSP was later generalized in~\cite{Partala_2008} for
left conjugacy closed (LCC) loops as a partial conjugacy search problem (PSCP)
allowing a wider class of platform structures.

In this paper, we show that cryptographic protocols that rely on infeasibility of
the CSP -- or, as is in general the case, infeasibility of
being able to conjugate with a secret element --
actually rely on infeasibility of the PCSP in a rack
(a left distributive left quasigroup). The rack is
induced by conjugations of left translations of the underlying structure.
Furthermore, the left translations of the rack retain the same conjugation structure
for its left translations.
This means that if an adversary can solve the PCSP in the induced rack, then
she is able to conjugate with any element of the original structure.
We suggest that any CSP based protocol could be implemented simply using a rack.
The binary operation could be induced by group conjugation, conjugation
of left translations of an LCC loop or by a completely different left distributive binary operation.
We give an exposition of the AAG protocol using a rack.
The protocol can be seen as a further generalization of~\cite{Partala_2008}.

\section{Preliminaries}
\label{Preliminaries}

Let $Q$ be a non-empty set with a binary operation $* : Q \times Q \to Q$.
We call $Q$ together with $*$ a \emph{magma} and denote it by $Q(*)$.
A mapping $L_{a}^{*}(x) = a * x$, where $a,x \in Q$,
is called a left translation by $a$.
We denote the set of all left translations of $Q(*)$
by $L_Q^*$.
If $L_a^*$ is a bijection for every $a \in Q$, then $Q(*)$ is a left quasigroup.
If there is no ambiguity about the binary operation, we leave it out
and write simply $L_a, L_Q$ and $Q$.

We denote the application of a left translation $L_a$ to an element $x$ by $xL_a$.
In this case, function compositions are worked out from left to
right. That is, for example, $xL_aL_b = L_b(L_a(x))$.
If $Q$ is a left quasigroup, then the left multiplication group of $Q$, $\LMlt = \left< L_x : x \in Q \right>$,
is the permutation group generated by the left translations.
A left quasigroup $Q(*)$ is \emph{left distributive} if 
\[
a*(b*c) = (a*b)*(a*c)
\]
for every $a,b,c \in Q$. It is \emph{idempotent} if $a*a = a$
for every $a \in Q$.
A left distributive left quasigroup is called a \emph{rack}~\cite{Joyce_1982,Joyce_1982_2,Fenn_1992}.
If a rack
is also idempotent, then it is called a \emph{quandle}.
An excellent survey of racks can be found in~\cite{Stanovsky_2004}.

A left quasigroup $Q$ is \emph{left conjugacy closed} (LCC) if
the set of left translations is closed under conjugation. That is, if
for every $a,b \in Q$ there are $c, d \in Q$ such that
\begin{equation}
\label{left_CC}
L_a^{-1} L_b L_a = L_c
\end{equation}
and $L_a L_b L_a^{-1} = L_d$.
A rack $Q$ is always LCC, since
\[
x L_a^{-1} L_b L_a = a(b(x L_a^{-1})) = (ab)(a (x L_a^{-1})) = (ab)(x L_a^{-1} L_a) = (ab)x = x L_{ab}
\]
for every $a,b \in Q$.

Let $G$ be a group and let $b, c \in G$ be conjugate. Given $b$ and $c$,
the conjugacy search problem (CSP)
is to find an element $a$
such that
\begin{equation}
\label{groupCSP}
a^{-1} b a = c.
\end{equation}
If $Q$ is a left quasigroup, then (\ref{groupCSP}) is not meaningful,
but we can consider the CSP in the left multiplication group.
In this case, given conjugate permutations $\beta,\gamma \in \LMlt$,
the problem is to find an element $\alpha \in \LMlt$,
such that $\alpha^{-1} \beta \alpha = \gamma$.
If $Q$ is LCC, it is useful to restrict ourselves to the case
$\beta = L_b, \gamma = L_c$.
Given $b,c \in Q$,
the problem is
to find $\alpha$, a composition of left translations and their inverses, such that
\[
\alpha^{-1} L_b \alpha = L_c.
\]
This is a partial version of the CSP (PCSP), originally described in~\cite{Partala_2008}
for LCC loops. (In~\cite{Partala_2008}, $\alpha$ was required to be a composition of
left translations, but this is only a slight generalization.)

\section{Conjugacy search and racks}

In this section, we shall consider the conjugation structure of the left translations of an LCC left
quasigroup. In order that a conjugation by $L_a$ is unique, we need the following definition.
\begin{definition}
Let $Q$ be a LCC left quasigroup for which there exists a function $\lambda: Q \times Q \to Q$
such that
\[
L_a^{-1} L_b L_a = L_{\lambda(a,b)}
\]
for every $a,b \in Q$.
The magma $Q(\lambda)$, whose binary operation is given by $\lambda$,
is called the \emph{left conjugation magma} of $Q$.
\end{definition}
If there is no such a function, then
we say that $Q$ does not have a left conjugation magma.
Structures for which such a function is defined include for example groups,
LCC loops and LCC left quasigroups that are \emph{effective}.
A left quasigroup is effective if the left translations are pair-wise distinct,
that is, if $L_a = L_b$ if and only if $a = b$.
\begin{proposition}
\label{left_conj_LDI_prop}
Let $Q$ be an effective LCC left quasigroup.
If $Q(\lambda)$ is the left conjugation magma of $Q$,
then $Q(\lambda)$ is a quandle.
\end{proposition}
\begin{proof}
$Q(\lambda)$ is a left quasigroup if and only if
$L_x^\lambda : Q \to Q$ is a bijection for every $x \in Q$.
We shall first show that $L_x^\lambda$ is injective.
Suppose that $\lambda(x,a) = \lambda(x,b)$. Now,
\[
L_x^{-1} L_a L_x = L_x^{-1} L_b L_x,
\]
from which $L_a = L_b$. Since the left translations are pairwise distinct,
$a = b$, and $L_x^\lambda$ is injective.

If $Q$ is finite, then $L_x^\lambda$ is a bijection. However, if $Q$ is infinite
it is not immediately clear that $L_x^\lambda$ is surjective. To prove this, we
observe that every left translation $L_x$ of $Q$
is an element of the symmetric group $\Sym{Q}$.
Conjugation by $L_x$ in $\Sym{Q}$,
\[
\sigma(\tau) = L_x^{-1} \tau L_x,
\]
for every $\tau \in \Sym{Q}$,
is an inner automorphism of $\Sym{Q}$.
By the left conjugacy closedness of $Q$,
\[
L_x^{-1} L_y L_x \in L_Q 
\]
for every $x,y \in Q$ and $\sigma(L_Q) \subseteq L_Q$.
Similarly, by LCC, $\sigma^{-1}(L_Q) \subseteq L_Q$ and
$\sigma(L_Q) = L_Q$.

We shall now prove that $Q(\lambda)$ is left distributive.
Let $a,x,y \in Q$. We can write $L_x = L_a L_{\lambda(a,x)} L_a^{-1}$ and
$L_y = L_a L_{\lambda(a,y)} L_a^{-1}$.
Now,
\begin{eqnarray}
L_{\lambda(x,y)} & = & L_x^{-1} L_y L_x = L_a L_{\lambda(a,x)}^{-1} L_a^{-1} L_a L_{\lambda(a,y)} L_a^{-1} L_a L_{\lambda(a,x)}
L_a^{-1}\nonumber\\
& = & L_a L_{\lambda(a,x)}^{-1} L_{\lambda(a,y)} L_{\lambda(a,x)} L_a^{-1}\nonumber\\
& = & L_a L_{\lambda(\lambda(a,x),\lambda(a,y))} L_a^{-1}.\nonumber
\end{eqnarray}
That is,
\[
L_{\lambda(a,\lambda(x,y))} = L_a^{-1} L_{\lambda(x,y)} L_a = L_{\lambda(\lambda(a,x),\lambda(a,y))},
\]
from which by pairwise distinctness of the left translations
\[
\lambda(a,\lambda(x,y)) = \lambda(\lambda(a,x),\lambda(a,y)),
\]
and $Q(\lambda)$ is left distributive.

In addition,
\[
L_x^{-1} L_x L_x = L_{\lambda(x,x)} = L_x
\]
for every $x \in Q$
and $Q(\lambda)$ is idempotent.
\end{proof}
\begin{proposition}
\label{conjstructure}
Let $Q$ be an effective LCC left quasigroup and let
\[
\alpha = L_{a_1}^{\epsilon_1} L_{a_2}^{\epsilon_2} \cdots L_{a_n}^{\epsilon_n},
\]
where $n \in \N$ and $a_i \in Q, \epsilon_i \in \{-1,1\}$ for every $i \in \{1,2,\ldots,n\}$.
If $Q(\lambda)$ is the left conjugation magma of $Q$, then
\[
\alpha^{-1} L_c \alpha = L_{c \alpha^{\lambda}},
\]
where $\alpha^{\lambda} = (L_{a_1}^{\lambda})^{\epsilon_1} (L_{a_2}^{\lambda})^{\epsilon_2} \cdots (L_{a_n}^{\lambda})^{\epsilon_n}$.
Furthermore,
\[
(\alpha^{\lambda})^{-1} L_c^{\lambda} \alpha^{\lambda} = L_{c \alpha^{\lambda}}^{\lambda}.
\]
in $Q(\lambda)$.
\end{proposition}
\begin{proof}
If $a_1 \in Q$, then $L_{a_1}^{-1} L_c L_{a_1} = L_{c L_{a_1}^{\lambda}}$ and
$L_{a_1} L_c L_{a_1}^{-1} = L_{c (L_{a_1}^{\lambda})^{-1}}$.
Since $Q(\lambda)$ is left distributive,
$(L_{a_1}^{\lambda})^{-1} L_c^\lambda L_{a_1}^\lambda = L_{c L_{a_1}^{\lambda}}^\lambda$
and
$L_{a_1}^{\lambda} L_c^\lambda (L_{a_1}^\lambda)^{-1} = L_{c (L_{a_1}^{\lambda})^{-1}}^\lambda$
The result follows from induction on $n$.
\end{proof}
By proposition \ref{conjstructure}, the left conjugation magma of $Q$ retains the conjugation structure
of the left translations of $Q$. Suppose that
$L_c$ and $L_d$ are conjugate in $\LMlt$.
Suppose also that it is feasible to solve
the PCSP in the left conjugation magma of $Q$.
This means that it is
feasible to find
$\alpha^{\lambda} \in \LMlt^{\lambda} = \left< L_x^{\lambda} : x \in Q \right>$
such that $(\alpha^{\lambda})^{-1} L_c^{\lambda} \alpha^{\lambda} = L_d^{\lambda}$.
By proposition~\ref{conjstructure},
\[
L_{x \alpha^{\lambda}} = \alpha^{-1} L_x \alpha
\]
and we are able to conjugate any left translation of $Q$ by $\alpha$
knowing $\alpha^{\lambda}$. This is enough to break cryptographic protocols that are based on infeasibility
of conjugating with a secret element. 
A necessary condition for the security of such
protocols is the infeasibility of solving the PCSP in the left conjugation magma.
In fact, protocols based on conjugacy search could be defined using a rack by
conjugating its left translations.
For example, if $Q$ is a rack, then the AAG protocol
can be implemented the following way.

Suppose that the participants are Alice and Bob.
Let
\[
S_A = \{ a_1, a_2, \ldots, a_s \}, \quad
S_B = \{ b_1, b_2, \ldots b_t \}
\]
be two publicly assigned subsets of a rack $Q$.
Let also
\[
\LMlt_A = \left< L_{a_1}, L_{a_2}, \ldots, L_{a_s} \right>, \quad
\LMlt_B = \left< L_{b_1}, L_{b_2}, \ldots, L_{b_t} \right>
\]
be the corresponding subgroups of $\LMlt$.
Alice and Bob choose secret elements
$\alpha \in \LMlt_A$ and $\beta \in \LMlt_B$, respectively,
by randomly multiplying
a finite number of generators
and their inverses.
Alice computes
\[
c_1 = b_1 \alpha, ~ c_2 = b_2 \alpha, ~ \ldots ~, ~ c_t = b_t \alpha
\]
and transmits $c_1, c_2, \ldots, c_t$ to Bob. Similarly, Bob
computes
\[
a_1 \beta, a_2 \beta, \ldots, a_s \beta
\]
and replies with the corresponding elements.
For every $1 \leq i \leq t$,
\begin{equation}
\label{LDIprob_equiv_PCSP}
c_i = b_i \alpha \quad \Longleftrightarrow \quad L_{c_i} = \alpha^{-1} L_{b_i} \alpha,
\end{equation}
and Alice and Bob are able to compute
$\beta^{-1} \alpha \beta$ and $\alpha^{-1} \beta \alpha$ (or, rather $\alpha^{-1} \beta^{-1} \alpha$), respectively.
The common secret key is $\alpha^{-1} \beta^{-1} \alpha \beta \in \LMlt$.
It has to be infeasible to compute $\alpha$ given $b_1, b_2, \ldots, b_t$
and $c_1, c_2, \ldots, c_t$. By (\ref{LDIprob_equiv_PCSP}),
this is equivalent to solving a system of conjugacy
equations of left translations in $Q$.

It should be noted that
the binary operation does not have to be induced by group conjugation.
Any left distributive operation with bijective left translations can be used.
For example, if $G$ is a group and $f$ is an involutory
automorphism of $G$, then $a * b = a f(a^{-1} b)$ defines a rack on $G$.
Similarly, if $e$ is a central element of $G$ and $a * b = a b^{-1} a e$,
then $G(*)$ is a rack.
Other constructions of left symmetric racks from groups can be found in~\cite{Stanovsky_2005}.
The platform structure does not need to be a group, however.
Some examples arising from different constructions can be found, for example,
in~\cite{Joyce_1982,Brieskorn_1988,Fenn_1992}.
Such racks possibly offer much harder partial conjugacy search problems than the racks
that appear as conjugation magmas of groups.

\section{Acknowledgements}

The author wishes to thank Markku Niemenmaa for valuable comments and
suggestions regarding the manuscript.
This research was supported by the following foundations: Finnish Foundation of Technology Promotion, the Nokia Foundation, Tauno Tönning Foundation, Walter Ahsltröm Foundation and The Finnish Foundation for Economic
and Technology Sciences -- KAUTE.

\bibliographystyle{abbrv}
\bibliography{biblio}

\end{document}